# Some Comments on the Tests of General Relativity.

M. Anyon and J. Dunning-Davies,

Department of Physics,

University of Hull,

Hull  HU6 7RX,

England.

Email:  j.dunning-davies@hull.ac.uk

**Abstract**

A brief outline of the history of the discrepancies within Newtonian mechanics at the end of the 19$^{th}$ Century is given. The framework of General Relativity is briefly described and the famous 'tests' of General Relativity (GR) are considered and alternative solutions are discussed, with particular attention concentrating on the advance of the perihelion of the planet Mercury. The implications for the claims of relativity are discussed, all with reference to both pre and post 1915 publications.

**Introduction**

The aim of this article is to address the notion that the well known classic 'tests of general relativity' - The Anomalous Precession of the Perihelion of the Planet Mercury, The Gravitational Red-shift and The Gravitational Deflection of Light-rays - may be explained and derived with no reference to Einstein's theory of general relativity. The intention is to provide evidence that these phenomena can be explained and interpreted without the framework of general relativity, by using both modern sources and publications, as well as work that was published prior to Einstein's work. By providing evidence that these phenomena could, and were explained, via classical mechanics before 1915, it will be possible to make a case against those that say anyone who questions general relativity is 'obviously' mistaken.

**Theories of the Relativity of Motion.**

The General Theory of Relativity (GR), proposed by Albert Einstein in 1915, was heralded as one of the most ground-breaking and complicated theories in physics when it was first published, and is still considered one of the greatest physical and mathematical theories of the universe at large today. The paper *'The Foundation of the General Theory of Relativity',* [1] was a more general derivation of his earlier work *'On the Electrodynamics of Moving Bodies',* known better as the Special Theory of Relativity (SR) (1905), and was essentially a theory of gravitation. The first paper was a consideration of the kinematical properties of uniformly moving inertial bodies, and was very similar to work by Lorentz, Poincaré, Minkowski and Larmor. It states, basically, that the laws of physics are invariant under a particular group of space-time co-ordinate transformations - namely the Lorentz transformations - and that no one system of co-ordinates may be considered as having special status over another. [2] The assumption that the laws of physics may be expressed in a form which is independent of the co-ordinate system is known as the *Principle of Covariance.* [3] This is a very important part of the theory, and the application of this principle means that the laws of nature will be the same for any reference frame and that no reference frame can be considered

'absolutely' at rest compared with others. The treatment of these principles of relativity by Einstein was similar in construction and mathematical basis to the theories and ideas being worked on by such as Lorentz and Poincaré, who had taken up the challenges of attempting to explain the last few remaining anomalies within Newtonian mechanics, including those posed by the subject of the relativity of motion.

The General Theory of Relativity, however, was something completely different. It was a new theoretical description of gravity that was based on a completely new mathematical framework and demanded a totally new way of thinking about the universe. Space could no longer be thought of as an empty volume with an aethereal travelling medium but as a four dimensional continuum. Space is described by the usual 3 dimensions of space, as well as the fourth dimension, time. This leads to a description of a four-dimensional *space-time,* which is capable of describing the physical universe. Gravity is not actually a force, in the conventional sense, but is the manifestation of the distortion of the flatness of the space-time continuum by gravitating massive bodies, such as planets and stars. The gravitational field can no longer be described simply by the Newtonian gravitational potential, $\phi$, but by the ten components of the relativistic metrical tensor, which will be considered in greater detail later. These new concepts of gravitation and kinematics, coupled with the relatively unheard of tensor mathematics of Minkowski and Riemann that GR uses, led to the common popular belief that General Relativity was an incomprehensible and abstract theory that only a handful of scientists could even understand, never mind use.

The idea of 'relativity' was not a new one, however, having been considered by physicists such as Lorentz and Poincaré, among others. Indeed, in many physics text books for either college or university students, e.g. [4], the story of Einstein's relativity almost invariably begins with an explanation of Michelson and Morley's failed experiment to detect the Earth's relative motion through the luminiferous aether. Most are then led to believe that Special Relativity, published 7 years after this experiment, was specifically developed as an explanation for this null result. Anybody who had heard of a 'relativity theory' in the late 19[th] and early 20[th] century would almost certainly have thought of Lorentz and Poincaré's theory of relativity. This was a theory which attempted to explain the kinematical and dynamical laws of physics with reference to sets of co-ordinate axes known as inertial reference frames, in much the same way as Einstein's SR paper did. Simply put, they were methods of describing the motion of bodies within space, or space-time, relative to each other. There was a significant difference between their respective theories however; Einstein removed the aether completely, whereas Lorentz and Poincaré retained it. This difference between the theories is what would cause the greatest debate, and controversy.

The aether is a concept that seems to have been fairly quickly abandoned from mainstream physics after the publication of Einstein's general relativity paper. Sometimes known as the 'Luminiferous Aether', this was a medium that was thought to pervade all space throughout the universe, providing a travelling medium for electromagnetic waves, such as light. The necessity of a medium for the propagation of light came from the discovery that light travels with a finite velocity. [5] This meant that action-at-a-distance laws were not allowable to correctly describe the propagation of light. Since it was understood, from the extensive works of James Clerk Maxwell, that electromagnetic waves oscillate in a medium and travel through the medium from source to observer, it was logical to propose that light propagation must occur with respect to a universal medium. Composed of miniscule particles that exist and pervade everywhere throughout the universe, this medium was called the 'aether'. Although there were, and still are, different aetherial theories, mainly differing on whether the aether behaves as a solid, gas, or a liquid, see for example [6], it was universally accepted that something existed as a propagation medium for light. Many experiments had been performed in an effort to detect the motion of the aether relative to Earth, with Michelson and Morley's being just one of them, but they had all failed.

But even before the end of the nineteenth century it had become clear to Poincaré that the failure of so many experiments to detect the relative motion of the Earth through the aether suggested a completely new possibility. In his lectures given at the Sorbonne in 1899 he described the experiments that had failed so far and explained his opinion.

Poincaré believed at this time that *'absolute motion is undetectable in principle'* [7] by any means. In the coming years he asserted his opinion to the physics community that he did not believe anything more than *relative* motions of material bodies could be detected and proposed a new principle into physics. This principle, which became known as the Principle of Relativity, was the impossibility of determining the velocity of the earth relative to the aether. In 1904 Lorentz, working separately, asserted the same general principle, [8] - cited in [7] and [9]. Working from this principle Poincaré declared; 'there must arise an entirely new kind of dynamics, which will be characterised above all by the rule, that no velocity can exceed the velocity of light.' [7] Given that Lorentz had already derived the equations required to make this new dynamics work, in the form of the Lorentz transformations, it was reasonably easy to develop a reformulation of physics obeying Poincaré's Principle of Relativity. By this reformulation it was found that: 'No inertial system of reference can be regarded as having a privileged status in the sense that it should be regarded as fixed while the others are moving: the notion of absolute fixity in space, which in the latter part of the nineteenth century was thought to be required by the theory of aether and electrons was shown in 1900-4 by the Poincaré-Lorentz theory of relativity to be without foundation.' [7] This principle, however, is exactly the same as Einstein's first postulate of special relativity, which will be considered shortly. The Poincaré-Lorentz theory, along with its transformations, was applied to Maxwell's equations of electrodynamics. If these equations could be proven to be invariant after applying the Lorentz transformation then the theory would stand, and indeed it did. By 1905, the same year Einstein published his first relativity paper, Poincaré had completed his theorem and had proved that the fundamental equations of the aether and electrodynamics remain unchanged after using the transformations.

So, by 1905, without the use of the four dimensional space-time, it had been shown how the null result of the Michelson-Morley experiment could be consistent with aether theory. The aether must exist in some form in order for electro-dynamical signals to travel, but it is impossible to detect our relative motion through it - by principle. Einstein's influence on this topic at this date was still limited. His 1905 paper did not gather a huge amount of attention upon publication and was considered little more than an abstract additional consideration to the Lorentz-Poincaré theory by most who had considered it.

The reason for this was that at face value, the difference between the Lorentz-Poincaré theory and the special theory of relativity was very limited. The Lorentz transformations were engrained within both the theories, and each gave as a principle that no velocity can exceed the speed of light, c, that it is impossible to detect the 'absolute motion' of a system through free space, or any aether-like medium which might be assumed to pervade it, and that the laws of physics are independent of the velocity of the co-ordinate system used. [3] Einstein's influence would not become significant until 1915, with the publication of GR, and the years following which led to the experimental verification of his theory.

**General Relativity**

With the publication of General Relativity in 1915, the differing views on the aether between the theories caused great debate and controversy over its fundamental basis. The general theory of relativity attempts to make full use of the general idea of the relativity of motion, so Einstein derived it as a generalisation of his SR paper. This meant that the laws of physics, including gravity, must be consistent with the above principles of special relativity. Debates arose, however, regarding the difficulty in reconciling Maxwell's theory of electromagnetism with Einstein's kinematical theories of relativity. The equations were the same as the Lorentz-Poincaré theory due to the use by both of the Lorentz transformations, but Maxwell's well established theory of electromagnetism demanded the existence of a universally pervading aether as a first principle, necessary for the propagation of electromagnetic waves. General relativity completely disregards the aether and postulates that it is meaningless to speak of absolute velocity through free-space. But this was not of great importance to Einstein, who merely wished to describe the universe, not explain it at its fundamental level. A distinction needed to be made between Maxwell's *theory* of electromagnetism, and Maxwell's *equations* of electromagnetism. The equations had been proven to be invariant under a Lorentz transformation by Lorentz and Poincaré so they could be maintained, but it would have been extremely unwise to suggest at the time that Maxwell's theory of electromagnetism was in error - the theory had become so revered. Einstein simply ignored the issue and continued with his predictions and calculations. It couldn't be denied that, even without the aether, Einstein's theory was proving itself extremely accurate and useful for describing the universe.

Many, though, found it difficult to reconcile their beliefs in the aether, ingrained in the minds of scientists by Maxwell, with the hugely successful predictions of GR. This is due to the abstract nature of the underlying concepts of general relativity. As stated previously, according to Einstein's theory, the universe is described by a four-dimensional space-time continuum., which many found hard to comprehend. When no mass is present the universe can be thought of as having a flat geometry since there are no distortions due to gravity. Newton's laws of motion state that in the absence of any external force the path of a particle in free space will be a straight line, or geodesic. The description of the path of an inertial body through flat space given by special relativity is expressed by the geodesics of the space-time metric. But the space-time metric is not always that given by special relativity. When mass is present, what we would think of as a gravitational field corresponds to a deviation of the space-time metric from the flat geometry of special relativity.[10] Einstein derived his theory with this in mind. This is why the equations of space-time within GR return to the flat geometry of special relativity, corresponding to the Newtonian description of space, at large radial distances from a mass source.

To analyse these complicated situations it is necessary to make use of the mathematics of tensors in order to collect together the complicated tidal forces experienced by a particle within a gravitational field. In empty space general relativity states that this tensor has 10 independent components, and a total of 20 if there is matter present, described by the energy momentum tensor. But tensors are also important due to the significance that Einstein gave to the *principle of equivalence.* This is the principle that began Einstein's line of thinking on gravity and general relativity. [11] The principle of equivalence dates back to Galileo at the end of the 16$^{th}$ century, and is associated with the famous alleged experiment of dropping two different sized rocks from the Leaning Tower of Pisa. Galileo's brilliant insight was that each of the rocks would fall, and impact the ground, at the same rate, and time, when air resistance is neglected. For a modern version of this experiment see NASA's video of astronauts dropping a hammer and a feather simultaneously on the surface of the Moon. This effect demonstrates the proportionality of the *inertial* and *gravitational mass* of a body. The principle of equivalence is so named because of the fact that a uniform gravitational field is equivalent to an acceleration, and can be transformed away by appropriate choice of co-ordinate system. [11] In order for this concept to hold it is important that no set of co-ordinates be considered 'preferential' - the *principle of covariance.* Having these principles as a basis for GR, Einstein could construct his theory. All of theses ideas and strange concepts, as complicated and often counter-intuitive as they are, did nevertheless lead to extremely accurate predictions of gravitational phenomena, and explanations of anomalies that had eluded Newtonian mechanics for a long time.

**The tests of general relativity.**

Obviously General Relativity was not simply accepted on the basis of its innovative concepts, but was actually able to explain previously unaccounted for effects, as well as predict other consequences of the curvature of space-time, such as the effects given within the famous '*Tests of General Relativity'*. In most standard texts on GR reference is made to the three original tests of the theory. The three crucial tests that have existed since the theory's publication are given [12] as:- 'The Gravitational Red-shift', 'The Gravitational Deflection of Light Rays' and the famous 'Advance of the Perihelion of the planet Mercury'. Other possible tests have been proposed over the years, many of them have still to be recognised, but sometimes considered as the fourth test of GR is the Radar Echo Delay proposed by I. Shapiro. This test makes use of recent technological advances in electronics and radar signalling, which make it experimentally possible to test the motion of light rays as a function of time to an accuracy necessary for testing Einstein's theory. [2] The time required for radar signals to travel to and from the inner planets is measured and any relativistic delay can be calculated. The treatment of this test will be limited in the following discussion.

The perihelion advance of Mercury had already been noticed well before 1916, and efforts to explain it using Newtonian mechanics had failed, but the other two tests were claimed as predicted effects of Einstein's curved space-time. When Einstein correctly predicted and explained these effects, and when they were confirmed observationally, they were officially interpreted as a direct consequence of the underlying concept of GR - namely the curvature of space-time about a gravitating mass. However, it has become evident that there exists a sizeable amount of evidence showing that these results of general relativity can be derived without recourse to Einstein's theory. If this is so, how can they be considered direct tests of it? The evidence for, and details of, a number of the alternative methods for explaining these effects is what will be considered now.

Considered the most important of these 'tests' was the successful prediction by GR of the anomalous advance of the perihelion of the planet Mercury - almost considered the 'jewel in the crown' of general relativity. This was mainly due to the extremely small magnitude of the discrepancy, only 43 arc seconds per century, and the fact that Newtonian mechanics had been quite unable to account for the motion for some time. A theory that could adequately explain this phenomenon had been a long time coming. The problem had proven extremely complex, due to the fact that the system involved - the solar system - was exactly that; highly complex. Newtonian gravitational theory says that the orbit of a single planet around the sun will be an ellipse, with the sun at one of the foci. In reality the orbits are more complicated than simple ellipses because each of the planets' orbits is perturbed by the gravitational effects of the others. Subsequently the elliptical orbits slowly rotate. Classical mechanics was able to account for most of these rotations, except that attributed to Mercury. This 'anomalous' precession of the perihelion was first observed in 1859 by French astronomer Le Verrier and his observations sparked off different avenues of research in an effort to account for the motion. [13] Some even suggested that there could be another planet, known as Vulcan, in the solar system and which was causing the perturbation,. This would not have seemed such a strange idea at the time since the same problem had arisen after William Herschel discovered Uranus in 1781. Uranus' orbit was calculated precisely, taking into consideration all of the gravitational effects of the other known planets. But Uranus' observed orbit would still not fit exactly to the agreed theoretical position. It was thought that another planet must be perturbing the orbit, and Neptune was discovered by Johann Galle on 23$^{rd}$ September 1846 almost exactly where it had been predicted. The new planet Vulcan was obviously never discovered, and whilst the continuing problem of the perihelion shift of Mercury was being investigated classically Einstein applied his relativity theory to what was known as the *Kepler problem* - the determination of the orbits of the planets round a central gravitating mass. [14] This was achieved after the long process of determining the 'field equations' of relativity. These gave Einstein a theory that was capable of describing paths of particles in a gravitational field generated by other masses. [14] Einstein knew that he had to build a theory that was capable of describing the complex nature of gravitation, and the universe, utilising his ideas of curved space-time, but he also had to create this theory with the characteristic that at large distances from mass sources the universe can be explained by Newtonian mechanics. Although Newtonian gravitational theory had remained largely unchallenged for approximately 250 years it is not completely accurate. It is an excellent first-order approximation but under certain circumstances it becomes less precise. These are the situations - within strong gravitational fields for example - when relativistic effects become important and GR can be used to gain superior results. So when Karl Schwarzchild published his solution to Einstein's field equations for a gravitating mass point in 1916 [15] it became possible to arrive at a value for the perihelion shift of Mercury via general relativity. The results were very much in line with observation, 43" per century, and General Relativity was regarded as a great success. It seemed apparent to all that the unique way that GR represented the universe in a four dimensional space-time was especially suited to solving problems of this sort. The Schwarzchild solution, in fact, was crucial for deriving solutions to both the perihelion shift and the gravitational deflection of light, as well as being the corner stone solution responsible for black hole theory as we know it, as will be explained shortly. The solution gives a detailed description of the space-time about a mass source and was ideal for Einstein's purposes, and is one of the most important and influential scientific papers concerning fundamental general relativity. The successful prediction of these effects soon made the power and influence of Einstein's theory, along with the extremely significant solution derived by Schwarzchild, apparent to the world. It soon became ingrained within public knowledge that this publication had revolutionised the way scientists think about the universe, and that it was all made so successful by the pioneering genius of one man - Albert Einstein.

**Alternative methods for addressing the tests.**

In some text books concerning General and Special relativity the theory is portrayed in such a light that it appears completely impossible that other avenues of research could produce the same results. GR is so highly considered amongst the scientific community that to suggest that its results could be derived in a different way is considered either laughable, or, sometimes, offensive. This is far from an exaggeration but seems ridiculous when it is considered that GR itself was a completely new and alternative way of explaining gravity on its publication. It can sometimes be difficult to see past the veneration of GR, and to view it in context. As explained the concept of the relativity of motion was one that had been considered already, and although GR was a completely new conceptual way of describing the universe, it utilised ideas that were not completely original, or completely attributable to Einstein. It is a common misconception, for example, that in order for his theory to work, and successfully predict observable phenomena, Einstein alone devised and based relativity on certain principles or postulates which were intrinsic to the theory. One of these postulates is that gravitational effects propagate at the limiting speed of light, $c$. This concept, however, had in fact been considered in detail previously by others in an attempt to solve the perihelion problem of Mercury classically. [16]

It is important to remember, at this point, the work of scientists in the field of classical post-Newtonian physics in the late $19^{th}$ and early $20^{th}$ centuries. This was a period of time before Einstein's publications when many, although not all, physicists working on the anomalies within Newtonian mechanics came to a realization; they recognized that 'action at a distance' concepts in Newtonian physics, whereby the effects of forces can be felt over very large distance instantaneously, should be replaced by a finite propagation speed in order that the equations of motion take into account the delay associated with this propagation velocity. [17] Physicists such as Weber and Tisserand, who managed to calculate a 14" per century result for the perihelion advance of Mercury, as well as others, had been using the idea of a retarded potential for some time and had been making excellent progress. [18] Several physicists had been proposing different gravitational potentials with finite propagation speeds mainly in an effort to explain the discrepancy in Mercury's orbital precession, and Paul Gerber (1898) was one of the first to suggest this velocity should be the speed of light. [13] This line of thought was all a consequence of the work of theoretical physicists such as Gauss and Weber who had been investigating modifications to the Coulomb inverse-square law. They introduced 'a velocity-dependant potential to represent the electromagnetic field, consistent with the finite propagation speed of changes in the field.' [13],[19] It was discovered that this type of law, when applied to the gravitational potential on a large scale, could predict perihelion advance of the planets of the order of magnitude of observations. Paul Gerber, a German schoolteacher, published his paper, entitled *'Space and time propagation of gravitation'*, in 1898 with a velocity-dependant potential that gave exactly the observed perihelion advance of Mercury, 43" per century, as well as accurate results for a number of the other planets. This was achieved by assuming that gravitation propagates at the speed of light. [16] Gerber's results obviously also matched those given by Einstein derived from his general theory of relativity 18 years later. [1],[20] From his paper it almost seems that the purpose of Gerber's work was to demonstrate that gravity propagated at the speed of light, rather than to give a retarded potential that could predict the perihelion motion of Mercury. Indeed Gerber may not have fully grasped the significance of what he had proposed. His treatment of the problem, and physical analysis, were not completely free from error, as discussed in [19], but he had at least derived the correct answer where so many before him had failed. Regardless, though, of why Gerber's solution has gone relatively unheard of in the public domain, the question still stands as to whether the advance of the perihelion of Mercury result can be considered a *crucial* test of relativity. If alternative solutions are available, such as that given by Gerber, then it cannot be considered intrinsic to general relativity, and neither can the postulate that gravity propagates at the speed of light. In fact, a recent paper by Juame Giné [17] has shown that the values of the anomalous precession of Mercury's perihelion, as well as the gravitational deflection of light as given by GR, can be reproduced by using a retarded potential similar to that given by Gerber in 1898. A similar Newtonian explanation for the perihelion precession of Mercury has been advanced by Harold Aspden, and considered by Giné in [13]. Dr Aspden, in his *Theory of Gravitation* uses what he calls a 'retarded energy transfer' to analyse the problem of the perihelion advance in a classical manner [21], [22], [23]. Also, in '*Physics Unified*' [24] Aspden makes use of the result of Gerber as a starting point for a re-evaluation of the equations of motion. It is taken that if the 'normal assumption that energy propagation through empty space must be at the speed of light, c, we (can) define a retardation period T.' [18] This retardation term is to be thought of as describing the amount of gravitational energy in transit between two bodies, when propagating at the speed, c, and can be expressed as R/c, where R is the radial distance between the two bodies. The use and addition of this retarded energy term modifies Newton's Law of Gravitation from:

$$d^2u/d\varphi^2 + u = GM/h^2 \qquad (1)$$

to:

$$d^2u/d\varphi^2 + u = GM/h^2 + 3GM(u^2/c^2) \qquad (2)$$

which is the law of gravitation that emerges from Einstein's general theory of relativity, expressed in polar co-ordinates. This expression has been derived with no recourse to Einstein's work whatsoever, and is based solely on classical mechanics of the 19$^{th}$ century.

Harold Aspden has published much work on the subject of using classical mechanics' results to solve problems that are claimed as proofs or tests of general relativity. He proposes that physics could well have progressed quicker without Einstein's influence of relativity. He has suggested that re-introducing the concept of the aether to solve large scale problems could help towards gaining a truer understanding of how the universe works, without any referral to GR - which did away with the concept of an aether [21], [18], [25]. Poincaré and Lorentz, however, did include an aether within their theory of relativity, so this proposal is far from being *ad hoc*. In fact Dr Aspden believes that the reason the physics community has been unsuccessful in its quest for the 'Unified Field Theory' is due to the inhibiting influence of Einstein's four-dimensional space-time. Dr Aspden believes he has discovered this 'Holy Grail' of physics by taking a different approach to the problem than most. To quote his *Physics without Einstein*: 'the assumption has been that a Unified Field Theory will connect gravitation and electromagnetic theory, whereas the argument followed above (by Aspden in [18]) has been based on a field energy distribution that is the same as that of the electrostatic action.' [18] A controversial claim to swallow for many indeed, especially as Dr Aspden claims it demands the existence of an aether, but one which may well have a solid foundation. The previously mentioned result of 14 arc seconds for the perihelion advance of Mercury as derived by Tisserand in 1872 was based on the analysis of retardation effects at the speed of light in electrodynamics. It is argued in [18] that if Tisserand had properly considered the mathematical forms of energy potential and planetary motion, and used the correct conversion factor, the law of gravitation would have been modified in such a way as to give the correct 43" per century result. Aspden claims: 'Had this simple error in analysis been noticed back in 1872 the course of science history over the past hundred years would have been very different.' [18] This is both a fascinating argument, certainly one that would get the blood pumping on either side of the debate, and a refreshing perspective compared to the entrenched views of most of the physics community. Through researching similar approaches, it soon becomes apparent that a reasonable amount of work has been published in the area of solving the 'tests' of general relativity without appealing to Einstein's theory at all - either by returning to the classical laws of Newtonian mechanics, or by analysing the problem in a completely new light.

As stated, the same Schwarzchild solution of 1916 used to calculate the perihelion shift of Mercury also gave a calculation for the bending of light rays in a gravitational field. [14] This, the second test of general relativity, would be the subject of a determined investigative expedition sponsored by the Royal Society in 1919, and would eventually be the experimental results that led to Einstein's worldwide fame, The aim of the expedition in 1919 was to make use of the solar eclipse that was to occur on 29[th] May and to take measurements of the bending of light rays passing close to the sun, in an effort to confirm general relativity. By viewing stars in the vicinity of the Sun during a total eclipse it is possible to determine the change in apparent positions of those stars whose light has passed close to the edge of the Sun. The two joint expeditions to Sobral, Brazil, and the island of Principe, West Africa, sent to observe the 1919 solar eclipse, very nearly didn't happen at all though. With Einstein's paper having been published during the height of the First World War, there was no way to predict, in 1917 when the expeditions were planned, whether the war would permit such a journey. However, the expedition headed by Eddington reached Principe ahead of schedule and luckily the stormy weather broke in time for the team to make their observations. General relativity predicted that for a ray of light grazing the sun's limb the angle of deflection would equal to 1.75 arc seconds, twice that of the prediction of the same effect by Newtonian mechanics. [3] By the start of June 1919, when the calculations were completed, the results clearly showed that general relativity had correctly predicted the results. With this data, and that obtained in 1922 by Campbell and Trumpler, the General Theory of Relativity became the most famous scientific theory the world had ever known. It seemed that Albert Einstein, and his revolutionary ideas of curved space-time and relative inertial reference frames, had solved the mysteries of physics that had been bewildering scientists for so long.

The effect itself is predicted by GR by use of the Schwarzchild solution mentioned above. This solution is able to give the geodesic paths of particles within a gravitational field, which can then be used to calculate the deflection of light rays by the field. The use of a retarded potential, similar to that employed by Gerber (1898) to predict the perihelion advance of Mercury, can, however, adequately account for this deflection of light rays, as discussed in Juame Giné's paper [17].

A thorough modern analysis of three tests of GR including the bending of light rays, as well as the fourth - namely the delay of radar sounding waves - is provided by Bernard Lavenda in [26]. This paper addresses the tests of GR in a completely different way. In the introduction it is said that 'sometimes new insight can be gained in looking at old results from a new perspective.' [26] It is shown by Lavenda that, through the use of celestial mechanics, three of the tests of GR - the deflection of light-rays, the perihelion shift and time delay in radar sounding - can be treated as diffraction phenomena on the basis of Fermat's principle and the modification of the phase of a Bessel function in the short-wavelength limit. [27] It is noted that free particles follow geodesics in the curved space-time of a gravitational field, and that it is required to provide a way of distinguishing between the centrifugal and gravitational fields which cause acceleration. To achieve this it is assumed that the trajectory of a light ray in a static gravitational field can be determined in the same way as in an inhomogeneous refractive medium. Applying Fermat's principle - the principle that the path taken by a light ray between any two points is always the path of shortest time - to a flat metric 'yields the phase of a Bessel function in the periodic domain for a constant index of refraction', making it possible to determine the form of the trajectory of the light ray. [26] Through a rigorous analysis it is shown that the effect of gravity is to make the medium optically more dense, effectively increasing its refractive index, in the vicinity of a massive gravitating body, leading to rays of light curving in the direction of the Sun. This effect, along with the perihelion shift, are explained on this basis. This is quite a different approach from explaining these phenomena than has been considered already, but is excellent in demonstrating the fact that these 'test's' can be addressed in various ways.

The third test, the gravitational red-shift in spectral lines, is probably one of the least renowned tests. This is based upon the dependence of the wavelength of light on the gravitational potential of its source, and had in fact been explained before general relativity, and so as stated by Whittaker: 'does not, properly speaking, constitute a test of [the theory] in contradistinction to other theories.' [7] However, it was considered a qualitative test when the theory was published and so must be considered here. This effect is based around the *Principle of Equivalence*. It was considered a test of GR because the effect was predicted by Einstein as a dependence of the wavelength of light on the gravitational potential of its source. It is noted that the actual, or 'proper', time period of successive light pulses from an atom must be independent of its location, with reference to the *principle of covariance*. From this it is possible to derive a ratio of the observed wave-lengths of light, corresponding to a given spectral line, that originate from the surface of a star, $\lambda_1$, and from a great distance from the star, $\lambda_2$.

$$\lambda + \delta\lambda/\lambda = 1 + m/r \qquad (3)$$

or

$$\Delta\nu/\nu = \Delta\varphi/c^2 \qquad (4)$$

[12],[3]. This effect has been tested both astronomically and terrestrially and the predicted results are accurate to within experimental errors. This would seem to show that GR can correctly predict and explain this effect, but in fact the well-known expression for the red-shift of spectral lines due to a gravitational field may be derived with no application of the theory of general relativity. The expression above is actually fully derivable from non-relativistic principles, and two derivations of the above ratio are given in Adler *et al.* [12] with no mention of Einstein's field equations, yet it is still considered a 'crucial test' of relativity. These alternative derivations are based solely on the principle of equivalence, or by simply considering the mass-energy relation $E = mc^2$ and the interpretation that light is composed of energetic quanta with an 'effective mass'. This is the argument given in [28] and gives exactly the same ratio as that stated above, showing that this expression has nothing specifically to do with general relativity. To derive this ratio it is simply a matter of deducing a photon's 'effective mass' given by

$$m = h\nu/c^2 \qquad (5)$$

The equation expressing conservation of kinetic and potential energy can then be written as

$$h\nu - GMm/r = h\nu - GMh\nu/rc^2 = \text{constant} \qquad (6)$$

This allows the well known gravitational red-shift formula to be derived. For this it should be noted that as $r \to \infty$, $\nu \to \nu_\infty$

$$h\nu_\infty = h\nu - GMh\nu/rc^2$$
$$\Rightarrow \nu_\infty - \nu/\nu = -GM/rc^2$$

or

$$\Delta\nu/\nu = \Delta\varphi/c^2 \qquad (7)$$

[28]. It soon becomes apparent that the popular belief that GR is the *only* theory capable

of correctly describing gravity and space on a cosmological scale is far from true. Newtonian mechanics still remains a close approximation in most first-order cases and can be used as such, and in other situations general relativity is used to predict and explain results with greater precision. But it is not, and should not be, considered the *only* port of call when requiring an explanation of gravitational effects. It has been shown by Gerber, Giné, Aspden and others that the use of a retarded potential, or energy transfer, term within the equations of motion can account for the classic tests of general relativity. Also, as the gravitational red-shift may be fully derived with reference only to the laws of conservation of energy, or the principle of equivalence, this has also been proven to be independent of GR.

**Conclusions.**

At no point here is the validity of Relativity itself being questioned - the complex beauty that the mathematics presents is easy to appreciate and the theory is readily seen a brilliant way to conceptualize and represent the universe - but the *crucial* 'tests of general relativity' have been given a status that they cannot uphold, due to the existence of the alternative explanations for them.

Relativity has certainly been one of the most influential theories of modern physics, and has pushed our understanding of the universe forwards exponentially. But it also seems just as likely that the relativity theory of Poincaré and Lorentz could have gained just as much support and success. The null result of the Michelson Morley experiment had been solved in such a way as to retain the aether concept, and the Lorentz transformations used were exactly the same as those in SR. The 'tests' of GR would not exist because they could have been solved by the finite propagation speed of gravitation as given by Gerber and others. The finite propagation speed of light, *c*, and the inability to detect the earth's motion through the aether would also be consistent with their theory, as declared by Poincaré in 1905. Work considered by Harold Aspden has shown that these solutions can be fully derived classically. All of this information is freely available in scientific journals, and is discussed at length in many text books but is often ignored. Indeed the prevailing view that Albert Einstein and his theories of relativity are the only valid theory of gravity in the modern day maintains its stranglehold on the world. Within the scientific community the issue is obviously more complex, but, somewhere along the way, the general populous has been convinced of something which is quite misleading. General relativity is a hugely successful theory, and can certainly appear to be a theory that only a select few can fully comprehend, but it is one of many different viewpoints on the universe at large and should be considered as such.

After investigating the proposal that the 'Tests of General Relativity' may be explained and predicted without any recourse to Einstein's theory, it has been found that this is very much the case. Many in the public domain are led to believe that Einstein and his theories of relativity are the final word on gravitational theory but, in reality, this is not the case. Einstein and his work have reached levels of reverence so much that they are held up to be automatically true, with anybody brave enough to question them being branded a scientific heretic. In fact even the words and opinions of the man who gave birth to general relativity are ignored in order to maintain the status quo of 'conventional wisdom', as phrased by Dunning-Davies [29]. It is now for others to reflect on this position and do so dispassionately with minds cleared of any previously held views which might prejudice a true examination of the position. If nothing else, science must be open minded when searching for answers to questions; if not, then truth will escape discovery!